\documentclass{article}

\usepackage{arxiv}

\usepackage[utf8]{inputenc} 
\usepackage[T1]{fontenc}    
\usepackage{hyperref}       
\usepackage{url}            
\usepackage{booktabs}       
\usepackage{amsfonts}       
\usepackage{nicefrac}       
\usepackage{microtype}      
\usepackage{lipsum}		
\usepackage{graphicx}
\usepackage[square,numbers]{natbib}
\usepackage{doi}
\usepackage{authblk}

\title{High-throughput nanopore fabrication and classification using FIB irradiation and automated pore edge analysis}

\author[1,+,*]{Michal Macha}
\author[1,2,+]{Sanjin Marion}
\author[4]{Mukesh Tripathi}
\author[1]{Mukeshchand Thakur}
\author[1]{Martina Lihter}
\author[4]{Andras Kis}
\author[3,*]{Alex Smolyanitsky}
\author[1,*]{Aleksandra Radenovic}

\affil[1]{Laboratory of Nanoscale Biology, Institute of Bioengineering, Ecole Polytechnique Federale de Lausanne (EPFL), Lausanne, Switzerland}
\affil[2]{imec, Kapeldreef 75, B-3001 Leuven, Belgium}
\affil[3]{National Institute of Standards and Technology, Applied Chemicals and Materials Division, 325 Broadway, Boulder CO 80305, USA}
\affil[4]{Laboratory of Nanoscale Electronics and Structures, Electrical Engineering Institute and Institute of Materials Science}

\affil[+]{These authors contributed equally to this work}
\affil[*]{Corresponding authors: michal.macha@epfl.ch, alex.smolyanitsky@nist.gov, aleksandra.radenovic@epfl.ch}

\renewcommand{\shorttitle}{}

\hypersetup{
pdftitle={A template for the arxiv style},
pdfsubject={q-bio.NC, q-bio.QM},
pdfauthor={Michal ~Macha, Sanjin ~Marion},
pdfkeywords={MoS2, FIB, nanopore, nanofluidics},
}

\begin{document}
\maketitle

\begin{abstract}
Large-area nanopore drilling is a major bottleneck in state-of-the-art nanoporous 2D membrane fabrication protocols. In addition, high-quality structural and statistical descriptions of as-fabricated porous membranes are key to predicting the corresponding membrane-wide permeation properties. In this work, we investigate Xe-ion focused ion beam as a tool for scalable, large-area nanopore fabrication on atomically thin, free-standing molybdenum disulphide. The presented irradiation protocol enables designing ultrathin membranes with tunable porosity and pore dimension, along with spatial uniformity across large-area substrates. Fabricated nanoporous membranes were characterized using scanning transmission electron microscopy imaging and the observed nanopore geometries were analyzed through a pore-edge detection script. We further demonstrated that the obtained structural and statistical data can be readily passed on to computational and analytical tools to predict the permeation properties at both individual pore and membrane-wide scales. As an example, membranes featuring angstrom-scale pores were investigated in terms of their emerging water and ion flow properties through extensive all-atom molecular dynamics simulations. We believe that the combination of experimental and analytical approaches presented here should yield accurate physics-based property estimates and thus potentially enable a true function-by-design approach to fabrication for applications such as osmotic power generation, desalination/filtration, as well as their strain-tunable versions.
\end{abstract}


\section{Introduction}\label{sec1}

Defect engineering is an important tool for enabling and exploring various physical phenomena at the nanoscale. From electronics~\cite{Jiang2019DefectFunctionalities} and spintronics~\cite{Ahn20202DDevices} to optical~\cite{Zhang2021Super-resolvedDichalcogenidesb,Caldwell2019PhotonicsNitride,Glushkov2021EngineeringWater} applications, the control over the material's crystal lattice structure is paramount to pushing the boundaries of modern technology~\cite{Tuller2011PointEngineering,Zhong2022TowardSensitivity,Lin2016DefectDichalcogenides}. Atomically thin materials are especially interesting, as they also promise to become the next-generation porous membranes for desalination~\cite{Ali2018, Thiruraman2020IonsZero}, gas filtration~\cite{Thiruraman2020GasAperturesc}, energy generation~\cite{Macha20192DGeneration}, optoelectronics~\cite{Xue2020Solid-stateSensors}, biosensing~\cite{Drndic2014SequencingPores,Heerema2016,Merchant2010,Feng2015}, and mechanosensing~\cite{Fang2018HighlyEthers,Fang2019MechanosensitiveMonolayersb,Marion2020TowardsMechanosensing,Fang2018ControlledOptimization} applications. In these areas, defect engineering is used to fabricate nanometer-scale defects or pores. With atomically thin and van der Waals materials such as graphene or molybdenum disulphide (MoS$_2$) as host membranes, the membrane thickness can also be controlled during fabrication. By tuning the size, pore-edge termination, and pore density, one possesses sensitive control over the membrane properties such as ion selectivity~\cite{Vlassiouk2008} and water permeability~\cite{Heiranian2015}, making the level of precision in controlling the geometry and composition of the fabricated defects (nanopores) a crucial fabrication component~\cite{Macha20192DGeneration}. 

Among the numerous available nanopore fabrication methods~\cite{Macha20192DGeneration,Danda2019Two-dimensionalTransport,Fried2021InFabrication}, the electron beam~\cite{Graf2019FabricationNanoporesb} and ion irradiation~\cite{Thiruraman2018} techniques offer a desirable compromise between the precision of fabricated pore geometries and scalability of the irradiated area~\cite{Thiruraman2018}. However, even upon achieving the finest parameter control, one has to employ a thorough pore characterization and statistical analysis to begin obtaining the full picture in terms of the nanopore properties and the resulting membrane-wide behaviour. Multidisciplinary efforts in combining nanopore fabrication protocols with computer vision image analysis~\cite{Chen2021ComputerApplications} are showing a great promise for future development of membrane technologies and study of nanoscale physics.

In this work, we investigate Xe-ion focused ion beam (FIB) as a tool for scalable, large-area nanopore fabrication on atomically thin, free-standing monolayer MoS$_2$. We analyze the irradiation parameters and their effect on the obtained nanopore sizes and porosity based on transmission electron microscope (TEM) and aberration-corrected scanning transmission electron microscopy (AC-STEM) imaging. We present an image analysis script that enables automatic nanopore edge identification and classification as well as statistical and geometrical pore analysis based on AC-STEM nanopore images and overcome challenges of conventional, manual image analysis. For the first time, we demonstrate an experimental method to study the membrane-wide pore edge analysis. Finally, we connect this data to atomistic molecular dynamics simulations to gain a detailed insight into the water- and ion-permeation properties of the resulting porous membranes. Our findings provide a new methodology to precisely control and analyze nanoporous membrane properties through a deeper understanding of transport phenomena occurring at the nanoscale. 

\section{Experimental Section}\label{sec2}

\begin{figure}[h!]
\centering
\includegraphics[width=1\linewidth]{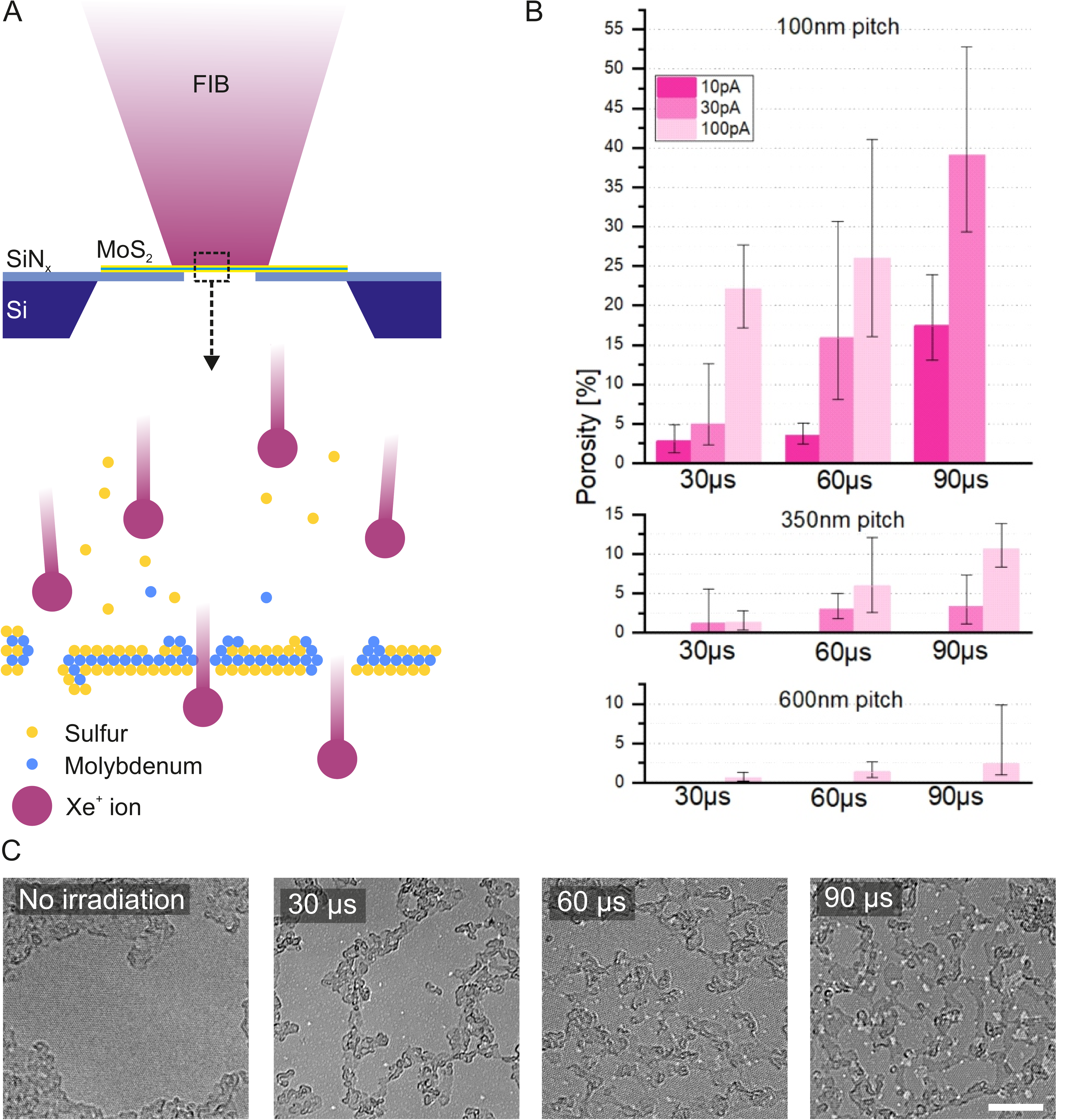}
\caption{\small \textbf{Focused ion beam irradiation.} A schematic representation of FIB irradiation principle (A) with Xe ions bombarding the free-standing MoS$_2$ to create nanopores. (B) Depending on process parameters such as beam current, pitch distance and dwell time formed nanopores vary in density and sizes.(C) Dependence on the beam dwell time on porosity and pore size. Scale bar- is 10nm.}
\label{fig:intro}
\end{figure}

\subsection{Focused Ion Beam Irradiation}
\begin{figure}[h!]
\centering
\includegraphics[width=1\linewidth]{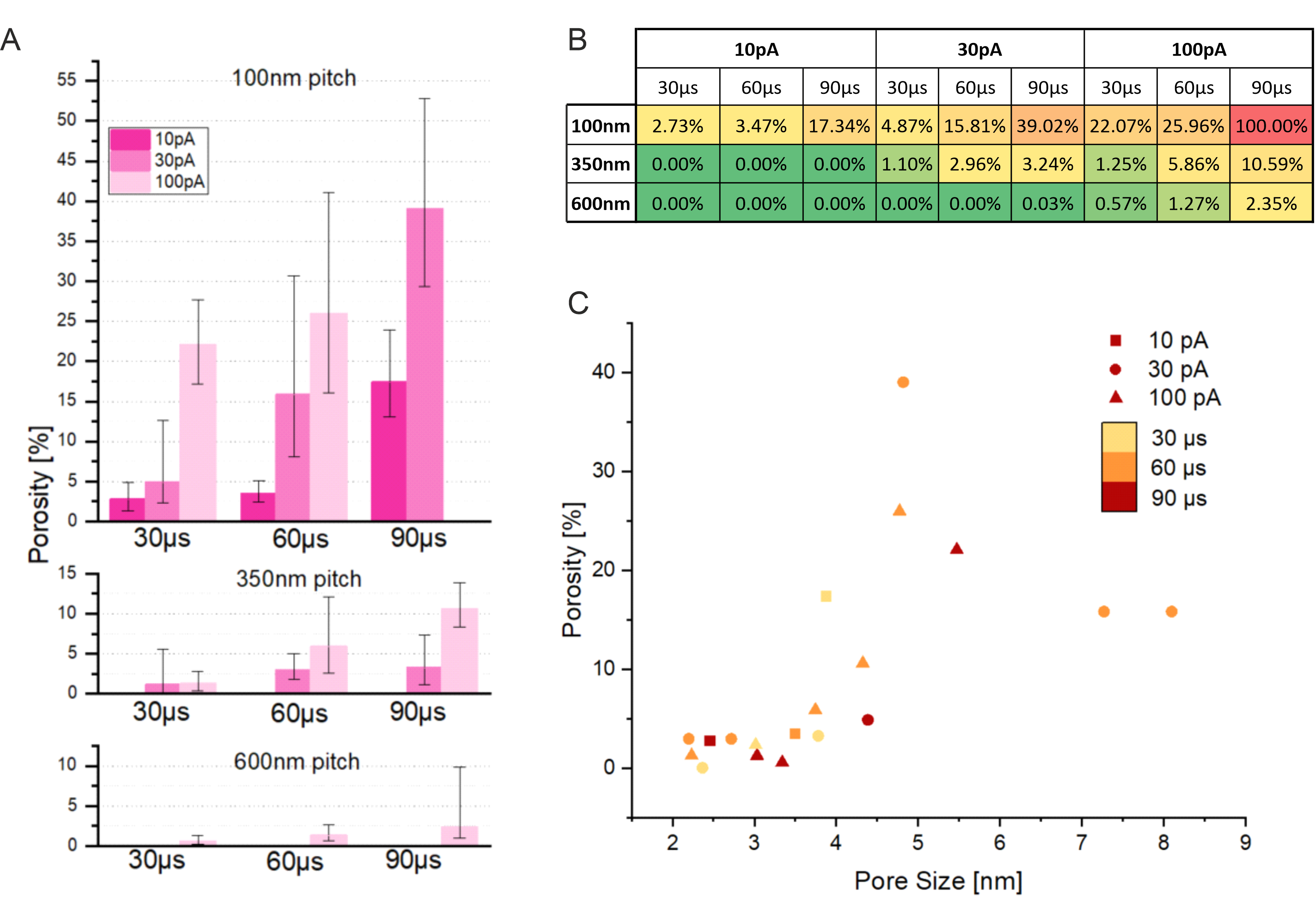}
\caption[Porosity and pore-size distribution]{\small \textbf{Porosity and pore-size distribution}. A dependence of membrane porosity versus used FIB parameters of dwell time, pitch-distance and beam current (A and B) for a constant 30kv Xe FIB. A  chart of achieved pore size (in the longest direction) and porosity is shown on panel C. These data were used to find and optimize nanofabrication protocols.}
\label{fig:SI-1}
\end{figure}

In our study, we used a Xe-ion plasma FIB system (see Methods section) as a nanopore fabrication platform. An MOCVD grown monolayer MoS$_2$ (see Methods section) was transferred over a Si/SiN aperture and irradiated with Xe ions (Fig \ref{fig:intro}A) under varied ion beam parameters to investigate the nanopore fabrication performance, \textit{i.e.} porosity and effective nanopore diameters (Fig. \ref{fig:intro}B,C). In all irradiation experiments, ion beam was kept at constant 30kV accelerating voltage with varying beam current of 10, 30, and 100pA, beam dwell time per pixel of 30, 60 and 90$\mu$s and x/y distances between pixels (pitch distance) of 100, 350 and 600nm. Ion fluence was kept relatively low between \(4.9\times10^{11}\) and \(1.7\times10^{14}\) ion/cm$^2$ to minimize beam damage and promote shower-like single ion collisions rather than wide-area milling. All irradiation experiments were performed with a single scan over the selected substrate area. Resulting membrane porosity was prescreened and quantified using Ilastik software package~\cite{Berg2019Ilastik:Analysis} (see Methods section) to obtain general insights into porosity and average pore size (Fig. S\ref{fig:SI-1}A). Within the investigated FIB parameters, the nanopore-yielding parametric window is relatively narrow. Large pitch distances tend to yield no nanopores at lower beam currents, while membrane breakage was observed at 100pA, 90$\mu$s, and 100nm pitch (Fig. S\ref{fig:SI-1}B), corresponding to \(1.7\times10^{14}\) ion/cm$^2$. The parameters and the effective diameters of the obtained nanopores are mapped in Fig. S\ref{fig:SI-1}C. A more detailed parameter description of the used beam current, dwell time, and pitch distance is accessible in the supplementary information (Fig. S\ref{fig:SI-2}-\ref{fig:SI-4}). After investigating a range of nanopore-yielding parameters through high-resolution TEM (HRTEM) imaging, a second, in-depth nanopore analysis was performed on a membrane irradiated with 30kV, 30pA ion beam, 350nm in x/y pixel pitch distance and 60$\mu$s dwell time per pixel (fluence of 6.77$ \times 10^{12}$ ion/cm$^2$). The porous membrane obtained with such a set of irradiation parameters yields pores of dimensions most relevant for applications such as desalination or biosensing (average of 1nm in Fig. S\ref{fig:SI-3}).

\subsection{Pore-edge detection} 

\begin{figure}[h!]
\centering
\includegraphics[width=1\linewidth]{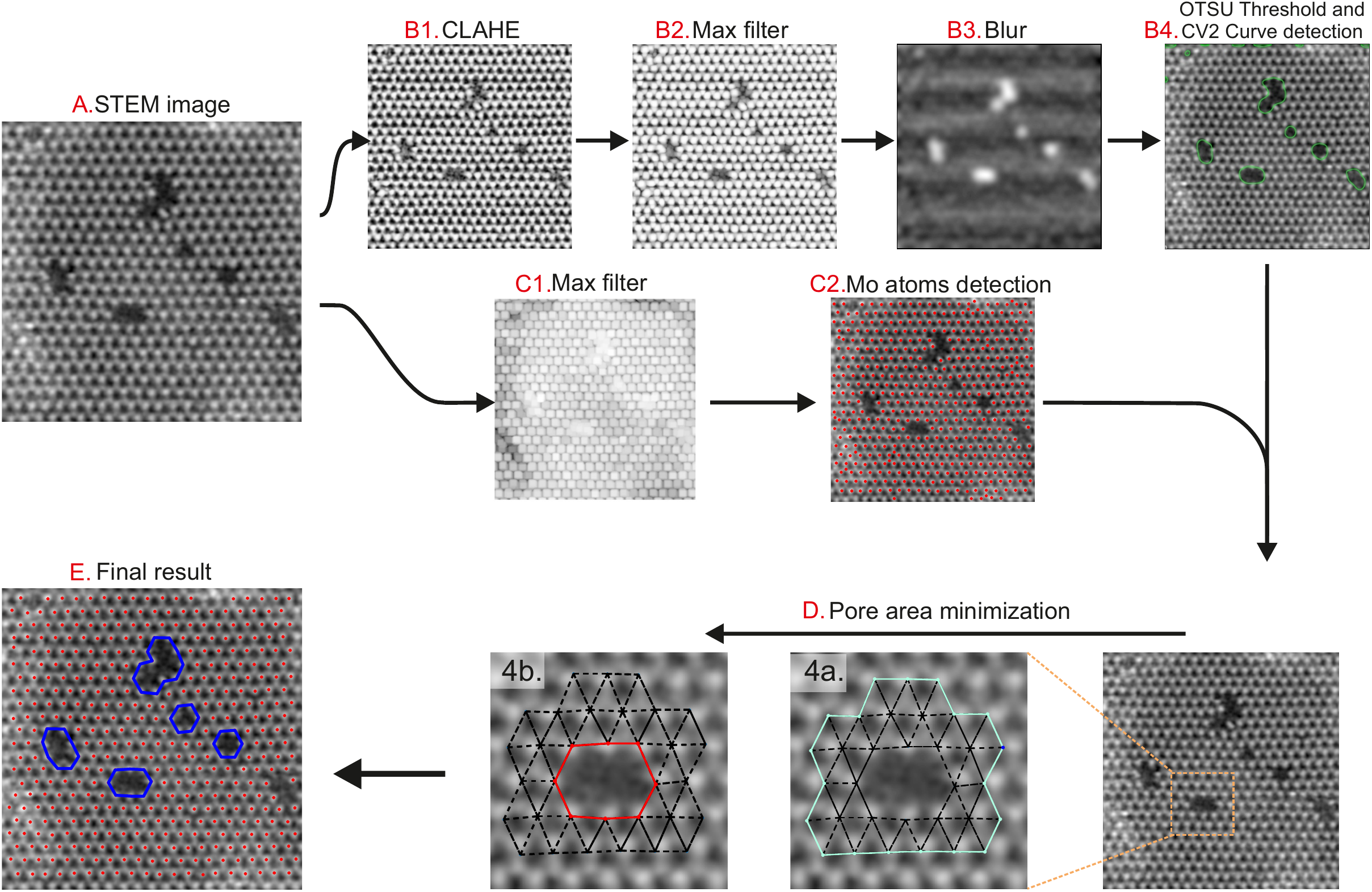}
\caption{\small \textbf{Pore edge detection algorithm outline.} A) The detection workflow starts with a STEM image, which is then processed in two parallel pathways. B) To obtain the locations of the nanopores first a CLAHE~\cite{Bradski2000TheLibrary} is performed (B1), followed by a maximum filter (B2) and blur of the inverse (B3). The resulting images are then Otsu thresholded, followed by the application of the OpenCV curve find routine~\cite{Bradski2000TheLibrary}, which provides the contours of the detected nanopores (B4). C) in the second pathway, a maximum filter (C1) is applied to the STEM image followed by a peak-finding routine (C2), which produces the location of the Mo atoms. D) Starting with each individual nanopore location from (B4), we take the Mo atoms surrounding the nanopore contour up to two nearest neighbours (4a), to which a custom optimisation algorithm is applied (4b), as described in the main text. E) The result of the pore detection algorithm constitutes the locations of Mo atoms that form the edge of the nanopore (marked as a blue polygon). }
\label{fig:alghoritm}
\end{figure}

\paragraph{AC-STEM image analysis} After prescreening the pore fabrication parameters with HRTEM imaging, MoS$_2$ irradiated at the fluence of 6.77$\times10^{12}$ ion/cm$^2$ was selected for further, in-depth analysis with an expected average pore size of $\approx$1nm. AC-STEM was used to take detailed images of the irradiated film (see Methods for details). The resulting images were then used for further analysis and automated pore edge detection (Fig.\ \ref{fig:alghoritm}). This task was divided into two parts: first, determining the positions of each atom species in the image, and second, determining the pore locations and extracting a polygon marking the pore edge. The latter also enabled us to study the distribution of fabricated pore shapes and the corresponding pore edge terminations.

\paragraph{STEM image processing and obtaining atom positions.} Obtaining the atomic positions in the lattice is critical to ensuring proper pore edge detection. A maximum filter with a constant fill is used with a lattice parameter roughly equal to approximately half the nearest-neighbour interatomic Mo-Mo distance $a_0$, \textit{i.e.}, a parameter value of $1/2a_0$. This maximum filtered image (Fig.\ \ref{fig:alghoritm}C is then processed to find the positions of the local maxima using the scikit-image~\cite{derWalt2014Scikit-image:Python} toolkit with a maximal separation of $3/8 a_0$. The resulting maxima correspond to the positions of the individual Mo atoms. The positions of S atomic sites can then be obtained by repeating this procedure with half the maximum filter lattice parameter used before, followed by removal of the already detected Mo atom peaks from the population. This approach is robust, as it depends exclusively on the lattice unit cell distance; it could be further improved by taking into account the lattice symmetries of the given 2D material. We find that the positions of the sulphur atoms are less reliable due to a lower intensity relative to that of the molybdenum atoms, as well as the various contaminants or dirt present on the imaged surface. Consequently, the Mo atom sites are used as the reference for defining the pore locations. 

\paragraph{Detecting the general pore locations.} Detecting pore locations using their general shape is a complex computer vision problem, which require the use of machine learning algorithms. In our case, due to the limited number of imaged pores and the complexity of the task, we opt for a classical image vision approach instead. We start with an approach similar to that described in Ref.\ \cite{Chen2021ComputerApplications}. We implement a contrast-limited adaptive histogram equalization (CLAHE) of the images to remove any variations in intensity due to variable illumination or dirt/contamination (Fig.\ \ref{fig:alghoritm}B1). Unlike in the previous studies~\cite{Chen2021ComputerApplications}, our aim is not to obtain the exact pore edge at this stage, but to identify prospective pore locations for further analysis. We first use a maximum filter and then blur the resulting images in order to remove any smaller areas of a different contrast, which enables us to emphasize the nanopore locations (Fig.\ \ref{fig:alghoritm}B2,B3). We use a maximum filter size, which is comparable to the Mo-Mo lattice distance $a_0$ in MoS$_2$. We then combine thresholding using the Otsu method~\cite{Otsu1979THRESHOLDHISTOGRAMS.} with OpenCV contour find~\cite{Bradski2000TheLibrary} to locate different pore contours (Fig.\ \ref{fig:alghoritm}B4). The Otsu threshold algorithm does not involve any user defined parameters, leaving the expected lattice parameter $a_0$ used in the atom detection and maximum filtering as the only user defined parameter in the analysis. The end result is an approximation of the nanopore edge shape, which is then used with the Mo atom positions to detect the exact pore edge.

\paragraph{Pore edge optimisation.} By using the obtained nanopore contours (Fig.\ \ref{fig:alghoritm}B4) and the positions of the Mo atoms (Fig.\ \ref{fig:alghoritm}C2), we are able to identify nanopore edges with reasonably high precision without freely adjustable parameters.  For each detected nanopore contour curve, we consider a group of Mo atoms consisting of any atoms within two typical Mo-Mo distances $2 a_0$ from the contour (Fig.\ \ref{fig:alghoritm}D). First, a convex hull is constructed from this group of atom positions, and then from it a polygon with edges between the sites corresponding to the distance between the nearest Mo-Mo neighbours ($\approx a_0$) is composed. This polygon (blue in Fig.\ \ref{fig:alghoritm} 4A) is then optimized using a custom algorithm (Supplemental Sec.\ \ref{section:SI-script}) so that its surface area is minimized, while maintaining nearest-neighbour distances on the triangular Mo sub-lattice. The algorithm seeks out any local changes in the polygon edges that reduce the total polygon surface area by removing or adding Mo sites. In case the algorithm is unable to converge to a valid pore shape, the group of atoms used for the original convex hull is expanded toward any nearby neighbouring groups of atoms, which did not yield a valid pore shape either. If no such group of atoms is found, the pore is then rejected from the statistic. This procedure automatically rejects all pores located close to the STEM image edge, such that parts of the missing edges are not visible. 

\begin{figure}[h!]
\centering
\includegraphics[width=1\linewidth]{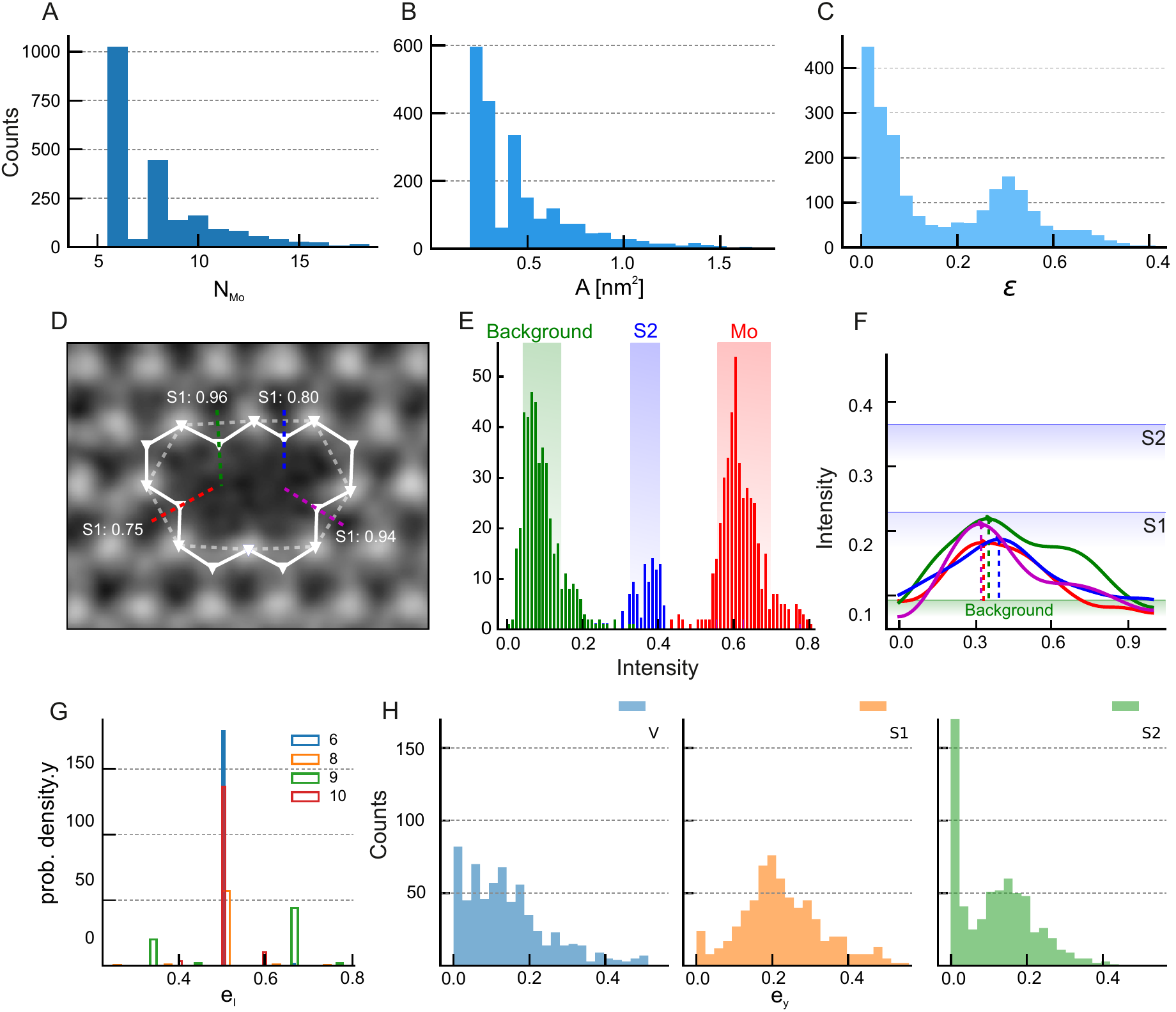}
\caption{\small \textbf{Pore edge analysis and statistical indicators obtained from pore edge detection.} A) A histogram of the number of Mo atoms detected in the population of nanopore edges detected using the custom-made edge detection software. B) Histogram of the area of detected nanopores. The reported area is that of the polygon marking the edge of the nanopore (crossing Mo and S atoms), C) A histogram of the elongation (eccentricity) parameter of the studied nanopore edge population. D) Magnification of a single nanopore with the detected nanopore edge (in white). Dashed lines of different colours indicate the location of intensity slices as shown in panel F. Letters indicate the detected pore edge atom type. E) Intensity histogram obtained from Mo and S$_2$ atom positions from one image. F) Intensity slices along the dashed lines of panel D. Horizontal lines mark the mean intensity level corresponding to various pore edge sites as determined from panel E. G) Normalized density of probability of finding a pore with a fraction $e_I$ of S$_2$ sites internal to the lattice (not exposed to the edge) for different numbers of Mo atoms present in the edge (6,8,9,10). H) Density of probability of finding a pore with a fraction $e_y$  sites populated by a vacancy (V), one sulphur atom (S$_1$) or two sulphur atoms (S$_2$) at the pore edge.}
\label{fig:edge-atom-detection}
\end{figure}

\paragraph{Identifying pore edges} Having obtained the coordinates of the Mo atoms forming a polygon outlining each nanopore, we can identify the composition of the pore interior edge and thus attempt to estimate its permeation properties on theoretical grounds. Depending on whether the edge composition is dominated by Mo and S atoms (carrying positive and negative partial atomic charges, respectively), a rich variety of tunable permeation properties is expected~\cite{Lee2014,Yang2017}, as also discussed in greater detail later in the text. In this sense, nanopores in single-layer MoS$_2$ are in fundamental contrast from those graphene-based, where pore edge termination appears to be close to neutral~\cite{Rollings2016} without functionalization. In addition, unbiased graphene membranes carry no overall surface charge density that could electrostatically boost osmotic currents through small pores. Within the scope of this work, we assume that the in-solvent MoS$_2$ pore structures are close to those imaged post-fabrication. Of course, further in-solvent functionalization (\textit{e.g.}, with hydroxyl groups) is possible to for example control hydrophilicity of the pore interior. More generally, the character of pore edge chemistry directly influences the electrostatic interactions inside the pore and thus further modifies the permeation properties~\cite{Surwade2015,Walker2017,Lee2014,Yang2017,Liu2018}. The first step in predicting the properties of porous 2D membranes is thus to identify the pore edge composition of the manufactured pores.

Starting with a polygon marking the Mo sites sites at the nanopore edge, we determine the positions of the sulphur sites in the lattice (Fig.\ \ref{fig:edge-atom-detection}). This subsequently enables us to identify the complete atomic pattern marking the edge of the nanopore. Such a pattern will have the form of Mo-$y_1$-Mo-$y_2$-Mo-$y_3$-\ldots with $y_k$
either a Sulphur site that is not directly exposed to the nanopore (it is interior to the lattice, marked as I) or an edge-exposed site. Exposed sites can be a double sulphur (marked as S$_2$), single sulphur (S$_1$) or empty (V). Intensities of atoms are known to scale with the atomic number~\cite{Krivanek2010Atom-by-atomMicroscopy} so we proceed by comparing the intensities at the edge atomic sites with reference values. As shown in Fig.\ \ref{fig:edge-atom-detection}, reference atom intensities are obtained from full STEM images, which usually feature multiple nanopores. The same peak identification algorithm as described in Fig.\ \ref{fig:alghoritm}C is used to detect S$_2$ and the background levels, except the lattice parameter is appropriately reduced and an inverse of the image used, respectively. The population of each sites is then obtained by removing each sub-population from the full statistic. The statistics for S$_2$ sites is of lower quality, as it is harder to identify positions of S2 in the lattice compared to the more intense Mo positions or similar S$_1$. The S$_1$ intensity value is determined as the average of the background intensity (V) and S$_2$ intensity values. Based on the obtained distributions of intensity values for different atoms, we proceed to identify each S site at the pore edge.
For each sulphur site at the pore edge we perform slicing of the intensity perpendicular to the sites corresponding to the Mo-Mo polygon edge (Fig.\ \ref{fig:edge-atom-detection}b), starting from the position of the dark spot and towards the nanopore interior. We calculate the mean intensity values and the corresponding standard deviations for each intensity group corresponding to four possible atom types (V, S$_1$, S$_2$, Mo). If a site does not have a prominent well-defined peak or is buried under the noise level defined by the background intensity distribution (V in Fig.\ \ref{fig:edge-atom-detection}a), it is identified as invalid (marked X). We find that invalid sites occur when either the pore edge was improperly recognized, or when the pore is contaminated. Full details of the algorithm are provided in the supplement. Assuming that the site intensity is obtained from the four independent normal probability distributions, we can extract the probability $p_k^y $ that a site $y_k$ on the edge corresponds to any of the three possible site types (V, S$_1$, S$_2$). 

We are now able to compare the different parameters obtained with analysis software from STEM images of irradiated samples. Fig.\ \ref{fig:edge-atom-detection}a,b,c shows histograms on the full statistic of nanopores identified using our algorithm. The distribution of the number of molybdenum atoms $N_{Mo}$  in the nanopore edge indicates that there are preferential shapes for the pores involving $N_{Mo}=6,8,9, \ldots$ Mo atoms, along with the corresponding nanopore areas as seen for the histogram of pore area values $A$. The elongation of different pores is represented by the eccentricity $\epsilon = 1- m/M$ with $m$ and $M$ being the minor and major axii of the nanopore, respectively, as estimated from the dimensions of the corresponding minimum and maximum bounding rectangles. This approach yields a value of $\epsilon=0$ for perfectly circular nanopores, and $\epsilon\rightarrow 1$ for a pore of infinite length in one dimension. In our case, there is a population of nanopores of circular shapes ($\epsilon \approx 0$), and those with a defined elongation of about $\epsilon \approx 0.3$ an $\epsilon \approx 0.5$, matching the most populous elongated pore shapes (marked as 8 in Supplemental Fig.\ S\ref{fig:SI-5-zoo}).

Our analysis also allows investigating the sulphur content of the nanopore edge. We define $e_y$ as the fraction of $y$-sites for each pore edge, such that the fraction of interior (non-edge-exposed) sites is $e_I = N_I / N_{Mo}$ and the fraction of $y=$V, S$_1$, S$_2$ sites, as $e_y=\sum_k p_k^y / N_{Mo}$. Here $N_I$ is the number of interior sites for a given pore. For arbitrary pore shapes, as shown in Fig.\ \ref{fig:edge-atom-detection}g, half of the sulphur sites are on average interior to the lattice, with another half exposed at the pore edge. A small sub-population of nanopores  (linked to the shapes associated with $N_{Mo}=9$, for instance) has longer zig-zag edge segments. Depending on the orientation of the lattice, this group produces two types of pores with different fractions of edge-exposed S-sites, as shown in Fig.~\ref{fig:edge-atom-detection}h. We note that there appears to be a slightly higher probability of finding S$_1$ sites at these edge-exposed sites, compared to S$_2$ or V sites. Overall, the dominance of Mo-terminated edges is due to the relatively high knock-on energy threshold needed for the incident beam to displace the Mo atom~\cite{Zan2013} compared to that required for S atoms. Additionally, this threshold is further lowered for the edge atoms due to reduced coordination, thus making edge sulphurs (especially the single-sulphur sites) fundamentally less stable during irradiation~\cite{Zan2013} as compared to molybdenum.

\begin{figure}[h!]
\centering
\includegraphics[width=1\linewidth]{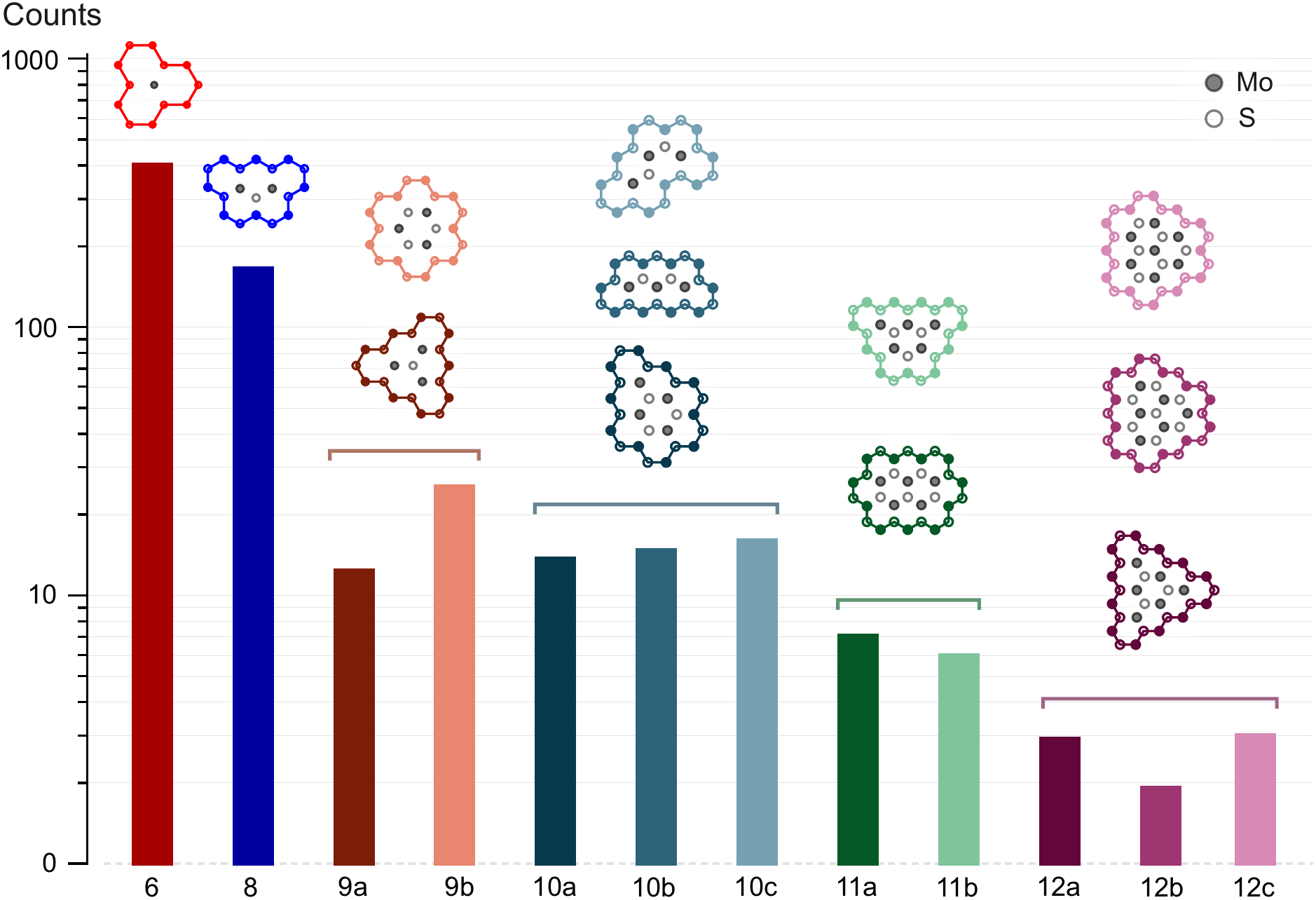}
\caption{\small \textbf{Pore edge shape distribution in Xe irradiated nanopores.} Histogram of identified pore shapes from STEM images of Xe irradiated MoS$_2$, here grouped in the clusters of pores containing 6, 8, 9, 10, 11 and 12 Mo atoms. Filled and open circles represent Mo and S atomic sites, respectively.}
\label{fig:pore-classy}
\end{figure}

\paragraph{Edge shape detection at the atomic resolution} Once the polygon shapes marking the Mo and S atoms at the nanopore edge are known, we can identify the likely nanopore shapes in the analysed images. Combinatorially, when the amount of Mo atoms lining the edge becomes large (in our case, $N_{Mo}>12)$, a near-exponential increase in the number of the pore structure possibilities is expected ~\cite{GovindRajan2019AddressingMaterials}. Taking into account that our analysis indicates a small probability of generating pores past that size, we limit our analysis of shapes up to $N=12$. This constitutes a maximum number of 3 (7) removed Mo (S) atoms from the lattice. We also limit ourselves to not considering the removal of pore edge exposed $S$ atoms, including them in our analysis of the site occupancy $n_y$ in Fig.\ \ref{fig:edge-atom-detection}h. To classify our pores (See Supplemental Fig.\ S\ref{fig:SI-5-zoo} for a complete list of pores included in our classification), we first remove from the statistic any nanopores which are severely deformed or contaminated (i.e. any edge S site is identified as invalid - X). Then we match the unique pattern of interior and exterior edge S atom sites between an unknown pore and the database of known nanopore shapes. In most cases misidentified shapes are due to the underlying atomic lattice being deformed, causing shapes which do not conform to normal lattice parameters. In our case, we note that about 1\% such invalid pores are identified for $N_{Mo}=6$, 6\% for $N_{Mo}=8$, 13\% for $N_{Mo}=9,10$, 27\% for $N_{Mo}=11$ and 34\% for $N_{Mo}=12$. If there are multiple shapes with the same pattern (in our case, for $N_{Mo}=12$ there are only two such shapes), we use the hu-moment based image recognition to determine the most probable match. Fig.\ \ref{fig:pore-classy} show the histogram of identified pore shapes up to N=12. Herein, all pores with invalid geometries, or with unidentifiable (classified as X) edge sites, have been removed from the presented statistic. We see that certain pore types are considerably more probable, \textit{e.g.}, the shape identified as 9b with more Mo atoms exposed to the pore interior is more probable than 9a, which has more sulphur sites exposed to the pore edge. Larger pores, with $N_{Mo}>10$, appear to be overall improbable in the Xe irradiated samples presented here. We find that the major source of error for the algorithm stems from the uncertainty in identifying what is an actual Mo atom in the lattice, and differentiating the atom from adatoms and contaminants. Most cases of irregular shapes identified with the algorithm, which are rejected from the analysis in Fig.\ \ref{fig:pore-classy} as they do not match a known shape, come from anomalous lattice atom positions. Such cases result from STEM imaging artifacts or unexpected adatoms or pore contamination. This can be improved upon in the future by using a better algorithm to detect atoms which takes into account the symmetry of the atomic lattice.

\begin{figure}[h!]
\centering
\includegraphics[width=1.0\linewidth]{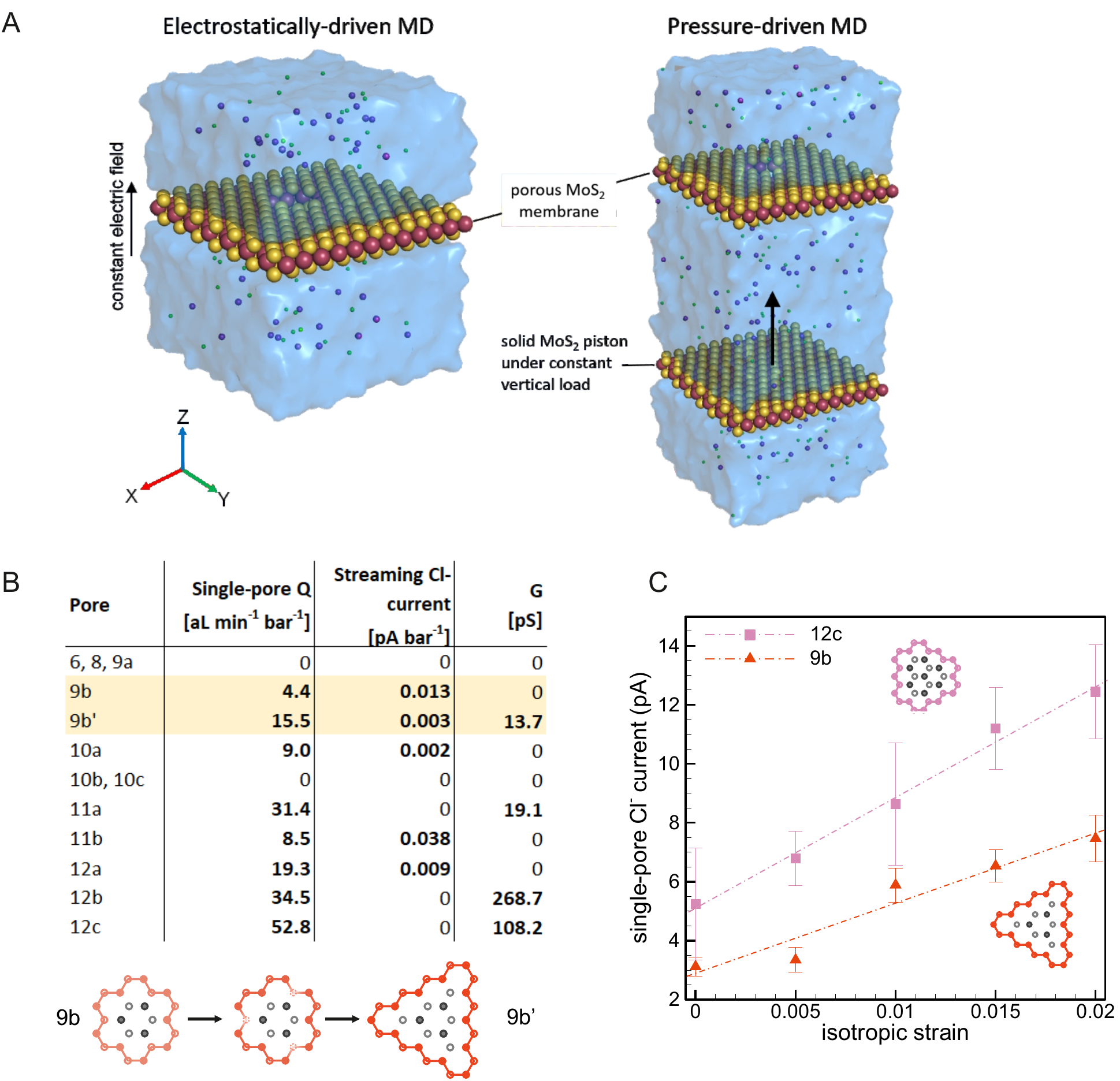}
\caption{\small \textbf{Simulated pore properties} A) Sketches of MD simulation setups for electrostatically- and pressure-driven systems. B) Ionic and water permeability properties for the identified pore shapes labeled as shown in Fig. \ref{fig:pore-classy}. Note that we classify pores 9b and 9b' as a specific cases with and without interior sulphur sites at the edge. C) Ionic current as a function of isotropic strain applied to membranes featuring select S$_2$- and Mo-lined pores. The bars in C are estimates of the average irregularities in the cumulative ionic fluxes used to calculate the ionic currents (\textit{e.g.,} see Ref. \cite{Fang2019MechanosensitiveMonolayersb}).}
\label{fig:pore-mechano}
\end{figure}

\paragraph{Permeation property prediction} 
Once the individual pore structures have been obtained, their properties can be estimated. Here we consider ion and water permeation through the pores comprising the statistics in Fig.\ \ref{fig:pore-classy} and shown in Fig.\ \ref{fig:pore-mechano}A, as obtained from all-atom molecular dynamics (MD) simulations described in detail in Section ~\ref{sec4}. To investigate ion flow driven by a constant electric field, MD simulations were set up similarly to our previous work on the transport mechanisms in subnanoporous 2D materials ~\cite{Fang2018HighlyEthers,Fang2019MechanosensitiveMonolayersb,BarabashFieldDependent}. For pressure-driven water flow, external hydraulic pressure was induced in the form of a constant force acting upon a solid piston. Sketches of systems driven by electric field and hydraulic pressure are shown in Fig. \ref{fig:pore-mechano}B. 

As summarized in Fig. \ref{fig:pore-mechano}B, most pores in Fig. \ref{fig:pore-classy} are ion-impermeable under a realistic electrostatic bias. The obvious factors here are the pore size (the number of removed atomic sites) and the pore eccentricity, \textit{i.e.}, sufficiently elongated pores preclude ionic permeability through high steric repulsion regardless of the pore length along the larger dimension. Other mechanisms consistent with earlier observations~\cite{Fang2019MechanosensitiveMonolayersb,BarabashFieldDependent} are more subtle and arise from the fact that Mo and S$_2$ carry partial positive and negative charges, respectively. Consider the fact that all ionic currents reported in Fig. \ref{fig:pore-mechano}B were contributed by anions (Cl$^-$). This observation likely owes to the electric field distribution near bulk MoS$_2$ monolayer surface, which is a finitely-spaced S$^-$-Mo$^{++}$-S$^-$ sandwich from the electrostatic standpoint. This electrostatic configuration appears to be overall repulsive to cations, also partly responsible for interlayer adhesion in multilayer MoS$_2$. Given this broad selectivity, anion transport is further boosted in pores featuring Mo$^+$-lined edges, despite the net zero charge of both the pore and the host membrane. Consider for example the 9b' and 12a pores, which are structural conjugates (geometrically identical, except the S$_2$ and Mo atomic sites are swapped): while 9b' lined with Mo atoms features sizeable ionic conductance, 12a is impermeable to anions. For the same reason, 12b, which is primarily lined with Mo atoms, has a significantly larger conductance than its structural conjugate 12c, except that now both 12b and 12c are ion-permeable due to their considerably larger size. 
Simulated pressure-induced water transport is shown to be rather low, being of order $ \frac{aL}{min \times bar}$, as shown in the table in Fig. \ref{fig:pore-mechano}B. Although such a low permeability renders the pore sizes considered here likely unfeasible for broad nanofluidic applications~\cite{Macha20192DGeneration}, given the overall high levels of ion rejection, they remain promising for desalination~\cite{Ali2018, Thiruraman2020IonsZero} and possibly gas filtration systems~\cite{Thiruraman2020GasAperturesc}.

The simulated properties of the individual pores identified here enable us to make several observations regarding mechanosensing~\cite{Fang2018HighlyEthers,Fang2019MechanosensitiveMonolayersb,Marion2020TowardsMechanosensing}, which is another emerging area of application for solid 2D nanopores. Some of the structures identified here have ion transport properties, which can be modulated by the application of mechanical strain to the membrane hosting the pore. Fig. ~\ref{fig:pore-mechano}C shows ionic currents under an electrical bias as a function of isotropic membrane strain. Observing this effect in experiments introduces a potential challenge. Currently, applying hydrostatic pressure in the direction perpendicular to the membrane appears to be the only realistic way to induce dilation of 2D nanopores~\cite{Davis2020Pressure-InducedNanoporesc}. Consequently, the resulting streaming flow may in principle compete with the effect of pore dilation on the ionic transport. Furthermore, depending on the relative directions of pressure and the external electrostatic bias, the two effects could in fact oppose each other. Although the data in Fig. \ref{fig:pore-mechano}B suggests overall low streaming couplings that at the most contribute currents of order 0.8 pA for pressures of order 20 bar, the coexistence of these effects must be taken into account when probing mechanosensitivity with the use of hydrostatically pressurized nanofluidic cells.

Once the individual pores have been quantified, for a given membrane hosting a large population of pores described in Fig. \ref{fig:pore-classy}, a predictive estimate of the membrane-wide permeation properties becomes possible. In the simplest case, assuming that the pores are spaced sufficiently sparsely so as not to affect each other's properties, these estimates are straightforward. For example, for a total of $N$ pores possessing a set of ionic conductance values $G_i$ and corresponding selectivity values $S_i$, the membrane-wide conductance and selectivity are $G_{m} = \sum G_i$ and $S_{m} = \frac{\sum G_i S_i}{\sum G_i}$, respectively. Similarly, the  mechanosensitivity (as defined earlier~\cite{Fang2018HighlyEthers}) for the entire membrane is $\mu_{m} = \frac{\sum G_i \mu_i}{\sum G_i}$, where $\mu_i$ are the individual mechanosensitivity values of the constituent pores. These estimates underscore the importance of obtaining accurate statistics of both the pore structures and the corresponding structure histograms. It is important to keep in mind that large numbers of impermeable pores in a given population have virtually no effect on permeability, while a relatively small number of highly permeable non-selective pores can readily "smear" the desirable membrane-wide properties. Overall, judicious use of the estimates discussed above may result in a promising fabrication parameter tuning tool that enables achieving desirable membrane-wide permeation properties and reduces the number of trial-and-error permeation measurements. 


\section{Conclusion}\label{sec3}

We have demonstrated the viability of Xe-ion irradiation as a large-scale nanopore fabrication tool. The presented irradiation protocol enables designing membranes with tunable porosity and pore dimension tuning, along with spatial uniformity across large-area substrates. Fabricated nanoporous membranes were characterized using STEM imaging and observed nanopore geometries were analyzed through a pore-edge detection script which depends only on one user defined parameter, the lattice constant of the 2D material in question. Our pore detection algorithm allows resolving individual pore structures and combining it into a  membrane-wide statistical analysis of the entire pore population. We then demonstrated that the obtained structural and statistical data can be readily passed on to computational and analytical tools to predict the permeation properties at both individual pore and membrane-wide scales. As an example, membranes featuring angstrom-scale pores were investigated in terms of their emerging fluid and ion flow properties through extensive all-atom MD simulations. 

High-quality structural and statistical descriptions of as-fabricated porous membranes are key to predicting the corresponding membrane-wide permeation properties. It is worth noting that as presented here and as would be the case with any image-based analysis, there fundamentally exists a degree of uncertainty in the resulting data. This arises from both uncertainty of the pore edge Sulphur sites being vacant, but also in the specifics of identifying the exact pore shapes. Therefore, the subsequent use of highly accurate (assuming proper parameterization), but computationally intensive methods (\textit{e.g.}, MD) to convert the structural and statistical data into permeation property predictions may be excessive. Instead, semi-analytical estimates of the permeation properties of individual nanopores and entire membranes could prove both reasonably accurate and highly computationally efficient. Although such estimation methods are challenging to implement, individual steps are already being taken towards for example rapid screening of nanopores ~\cite{screeningKarnik2021} or a detailed analytical formulation of how water molecules behave near nanopores ~\cite{Barabash2021OriginNanopores}. We believe that further improvements in the on-the-fly conversion of the structural and statistical data into an accurate physics-based property estimates should yield software tools that enable a true function-by-design approach to fabrication for applications such as mechanosensing, osmotic power generation, desalination, and filtration.

\section{Methods}\label{sec4}

\textbf{Substrate fabrication} MoS$_2$ was synthesised on c-cut sapphire wafer in metalorganic chemical vapor deposition (MOCVD) tube-furnace using large-scale growth methodology \cite{Thakur2020Wafer-ScaleMoS2b,Cun2019Wafer-scaleSiO2d}. As-grown material was transferred onto TEM SiN holey grids (Norcada) via polymer-assisted transfer method \cite{Thakur2020Wafer-ScaleMoS2b}. Before and after FIB processing suspended MoS$_2$ membranes were washed in acetone and soft-baked at 150$^{\circ}$C for 2h to minimize airborne hydrocarbon contamination and beam-induced deposition. 

\textbf{Focused Ion Beam Irradiation} 
Ion irradiation was performed at ThermoFischer Helios G4 PFIB DualBeam with Xenon Plasma FIB column. All FIB irradiation experiments were performed at constant 30kV with varying beam current, dwell time and pitch distance. Irradiated area was 200x200$/mu$m, centered at the substrate grid/membrane location.

\textbf{HRTEM image analysis and machine learning}
Transmission Electron Microscopy imaging was done on the ThermoFischer Talos F200S at 80kV accelerating voltage. Obtained high resolution images of nanoporous films were then cropped and bandpass-filtered. Subsequent machine-learning pore segmentation was performed in Ilastik and further processed and analyzed with Fiji ImageJ2. Due to the complexity of pore-detection and insufficient resolution, the more detailed analysis was performed with AC-STEM micrographs and custom-made analysis script described in the main text. 

\textbf{STEM imaging} 
Scanning transmission electron microscopy imaging was conducted using an aberration-corrected (with double spherical  corrector) FEI Titan Themis TEM 60 - 300 kV, equipped with Schottky X-FEG electron source and a monochromator to reduce the effect of chromatic aberrations. To avoid the electron-beam induced knock-on damage, a low acceleration voltage (80 kV) was used for the imaging \cite{Komsa2012Two-DimensionalDopingc}. The electron probe current, C$_2$ aperture size, and a beam convergence angle was 25pA, 50$\mu$m, and 21.2 mrad, respectively. Images were acquired with a Velox software (ThermoFisher Scientific) using 185 mm camera length which corresponds to an angular range (49.5 - 198 mrad) in a HAADF detector. To minimize the sample drift, a serial imaging was performed using 512 × 512 pixels with 8 $\mu$s dwell time. Images were aligned and processed using the \textit{double-Gaussian filtering} method in ImageJ.

\textbf{MD simulations} 
The all-atom MoS$_2$ model was set up according to recently published work~\cite{Sresht2017MoS2}, implemented within the OPLS-AA framework~\cite{Jorgensen1988,Jorgensen1996}. The atomic charges of Mo and S sites at the zigzag edges of the pores were set to $2/3$ of their bulk counterparts (for bulk charges, see Ref.~\cite{Sresht2017MoS2}), corresponding to $+0.3333$ and $-0.1667$, respectively. Such an assignment guarantees overall electrostatic neutrality of the zigzag edge and thus all pore structures considered in this work regardless of their geometry. Explicitly simulated water molecules were described according to the TIP4P model ~\cite{tip4p2004} while ions were described according to default OPLS-AA~\cite{Jorgensen1988,Jorgensen1996}. 

The simulations were performed using the GPU-accelerated Gromacs package~\cite{Van_Der_Spoel2005-sq,Hess2008} (version 2021.4). Ion and water transport through the pores was simulated in the presence of water-dissociated 1M KCl. The simulation cell sizes were approximately $3.82 \times 3.86 \times 5.0$ nm$^3$ and $3.82 \times 3.86 \times 10.0$ nm$^3$ for the ion and water flow simulations, respectively. In all simulations, the atoms at the perimeter of the porous membrane were harmonically restrained against displacement along X, Y, and Z ($k_X=k_Y=k_Z=34$ N/m), while the piston membrane in the pressure-driven simulations was free to move. The constant vertical force acting upon the piston was $F=pA$, where $p$ (in the range 10-50 bar) is the desired pressure, $A = 3.82 \times 3.86 = 14.75$ nm$^2$ is the in-plane cross-sectional area of the simulation cell. In the ion flow simulations, the external electric field (of order 0.02 V/nm) was $E=\Delta V/h$, where $\Delta V \approx$ $0.1$ V is the potential difference consistent with voltages typically applied in experiments and $h = 5$ nm is the height of the simulation cell, respectively. Periodic boundary conditions were applied in the $XYZ$-directions. After initial equilibration at constant $p$ = 1 atm and $T$ = 300 K, constant-volume production simulations were performed at $T$ = 300 K for at least 0.5 $\mu s$ to ensure reasonable permeation statistics.


\textbf{Authors Contribution}
M.M. synthesized MoS$_2$, developed the irradiation process and performed materials characterization and imaging; S.M. wrote pore-edge analysis script; M.Tripathi performed AC-STEM imaging; M.Thakur and M.L did material transfer; A.S. designed and performed MD simulations. M.M., S.M., and A.S. analyzed the data, developed the key concepts, and wrote the manuscript with inputs from all authors. A.R. supervised the project. All authors discussed the results and commented on the manuscript.

\textbf{Acknowledgments}
This work was in part financially supported by a Swiss National Science Foundation (SNSF) support through 200021$\_$192037 and the CCMX Materials Challenge grant “Large area growth of 2D materials for device integration”. A.S. gratefully acknowledges support from the Materials Genome Initiative.

\textbf{Disclaimer}
Commercial equipment, instruments, or materials are identified only in order to adequately specify certain procedures. In no case does such identification imply recommendation or endorsement by the National Institute of Standards and Technology, nor does it imply that the products identified are necessarily the best available for the purpose.

\section{Supplementary information}

\begin{figure}[h!]
\centering
\includegraphics[width=1\linewidth]{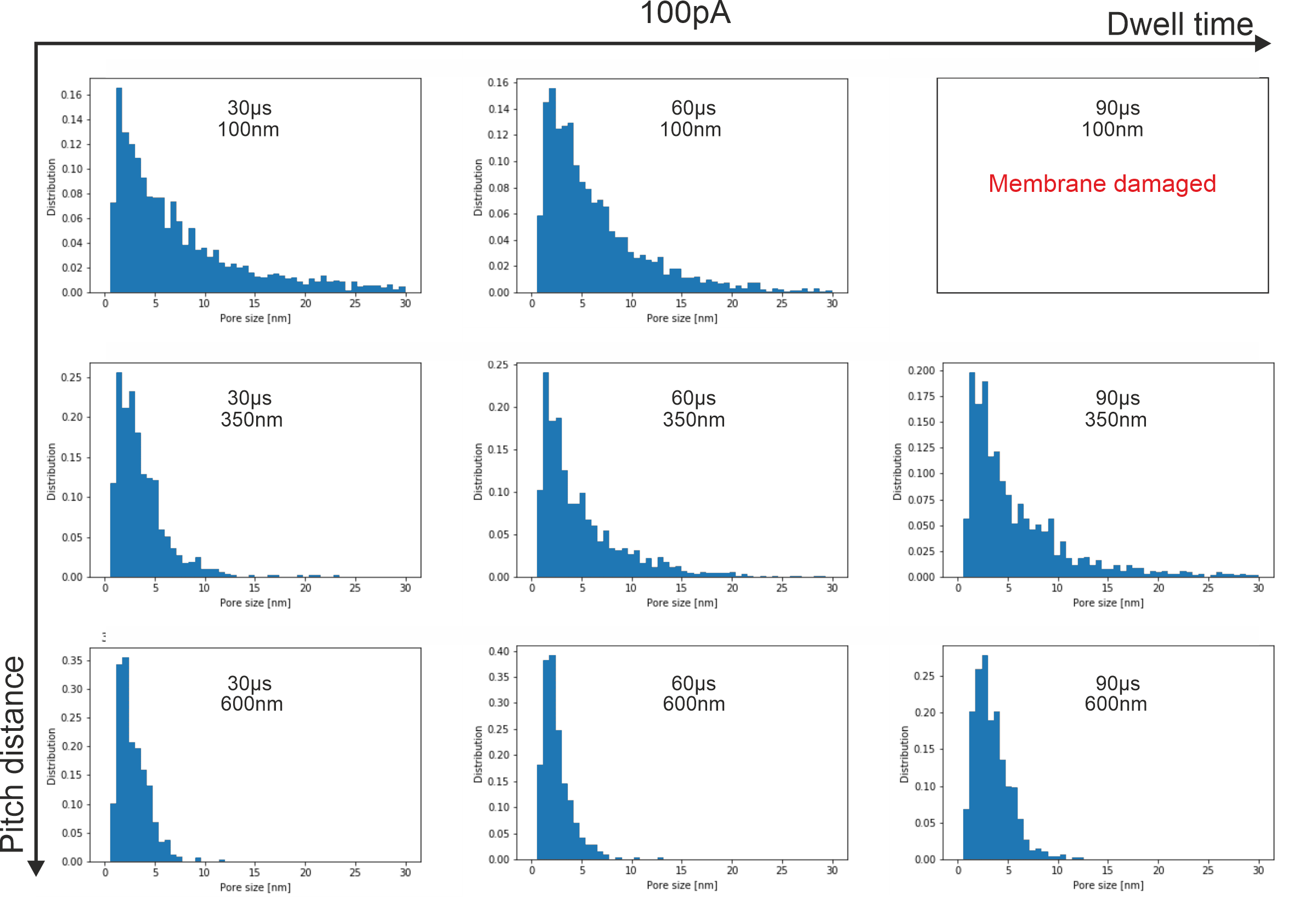}
\caption[Pore size histograms for 100pa FIB]{\small \textbf{Pore size histograms for 100pa FIB}. Histograms of pore sizes obtained at set irradiation parameters. Data obtained from the average of 10 HRTEM images taken from 10 substrates.}
\label{fig:SI-2}
\end{figure}

\begin{figure}[h!]
\centering
\includegraphics[width=1\linewidth]{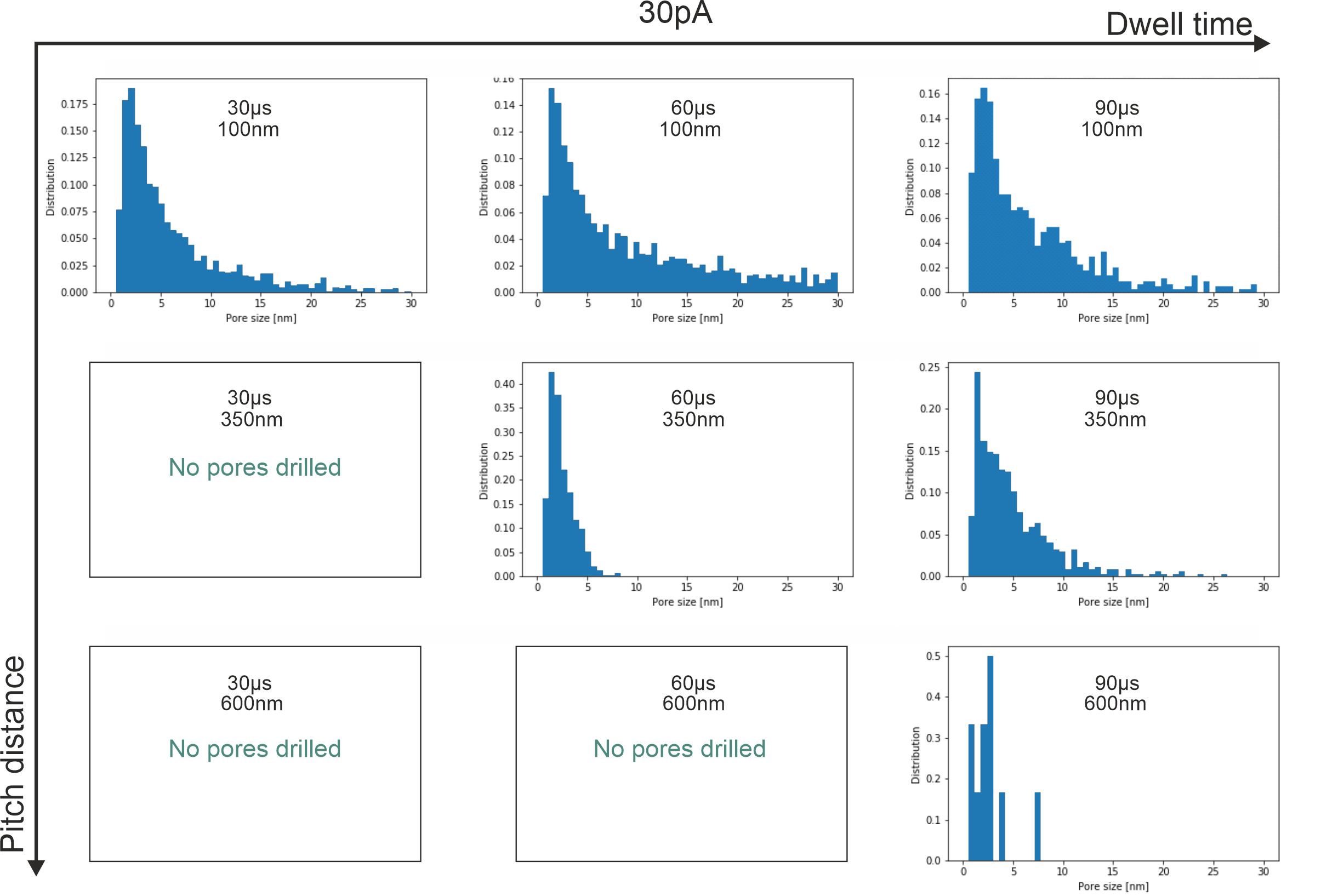}
\caption[Pore size histograms for 30pa FIB]{\small \textbf{Pore size histograms for 30pa FIB}. Histograms of pore sizes obtained at set irradiation parameters. Data obtained from the average of 10 HRTEM images taken from 10 substrates.}
\label{fig:SI-3}
\end{figure}

\begin{figure}[h!]
\centering
\includegraphics[width=1\linewidth]{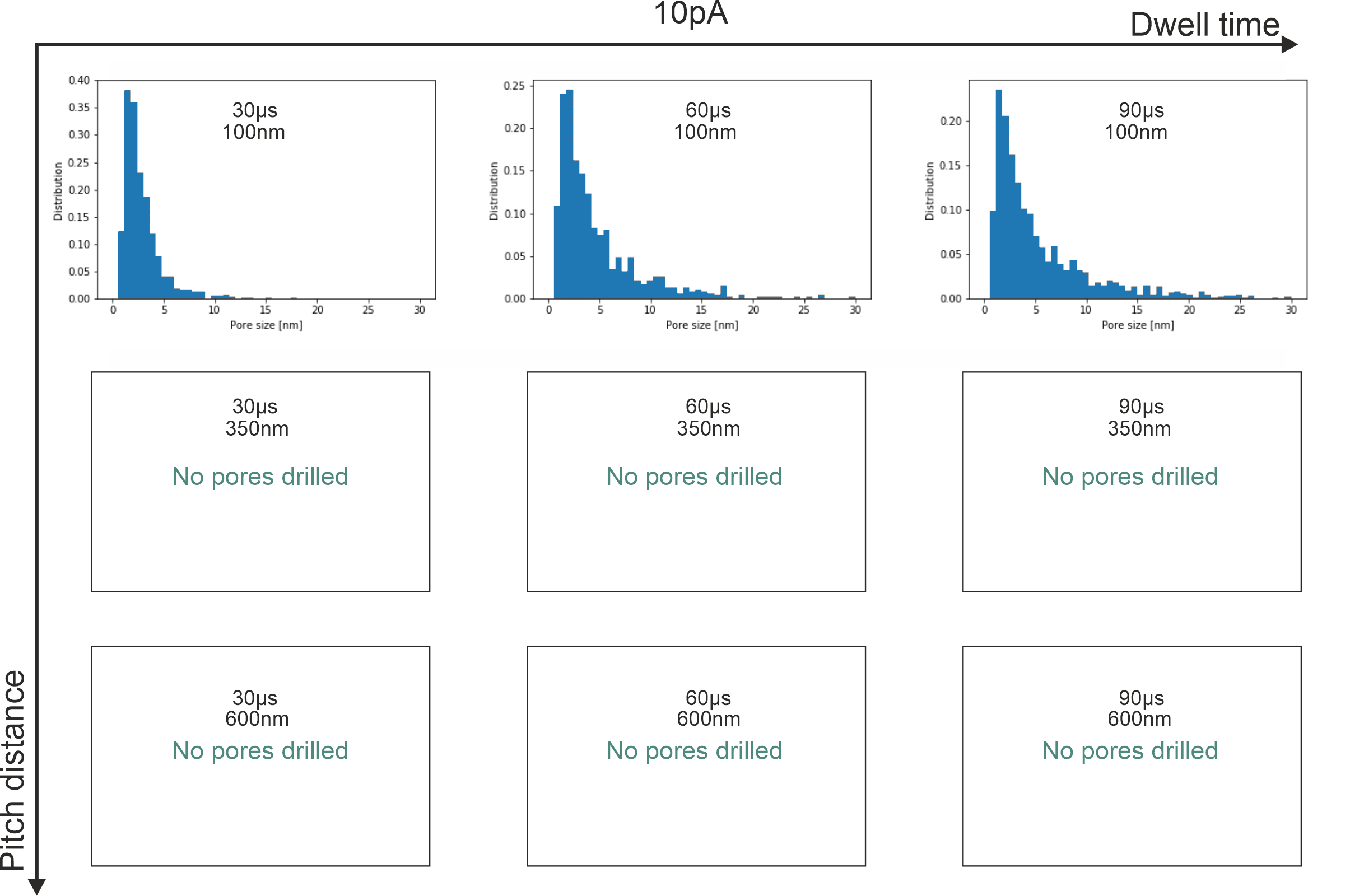}
\caption[Pore size histograms for 10pa FIB]{\small \textbf{Pore size histograms for 10pa FIB}. Histograms of pore sizes obtained at set irradiation parameters. Data obtained from the average of 10 HRTEM images taken from 10 substrates.}
\label{fig:SI-4}
\end{figure}
\clearpage

\subsection{Pore shape optimisation}
\label{section:SI-script}
To obtain the exact edge of a nanopore we start with a group of atoms $\{ a_i \}$ each with coordinates $\{x, y_i\}$ centered around a suspected nanopore location. For each pore we proceed as follows. 

First, a convex hull polygon is generated which by its definition envelops all the points in the group. For each edge pair of atoms in this polygon, we find the shortest path between them on a matrix of nearest neighbour atoms ($\delta_{i,j}$). A new polygon with $N$ points $P=\{p_k\}$ is then generated as a union of all shortest paths between the original convex hull polygon edge points. The resulting polygon approximates the original convex hull but with edge/side lengths corresponding to the individual Mo-Mo atom distances.

Secondly, we repeat the following procedure over the polygon, generating a new polygon $P'$ which then replaces the original polygon $P$ until the algorithm converges to the new polygon matching the old one ($\{p_k\}$ = $\{p_k'\}$). 

For each atom $p_i$ in the polygon $P$, taking into account that the polymer is cyclic ($p_{N+1} = p_1$), test for one of the following cases:
\begin{enumerate}
    \item \textbf{Remove dangling atom:} If $p_i$ - $p_{i+1}$ - $p_{i+2}$ form a dangling bond ($p_{i}=p_{i+2}$), then remove $p_i$ from the polygon $P$ to form $P'$ if the resulting polygon $P'$ would have a smaller area than $P$.
    \item \textbf{Remove extra atom:} If $p_i$  and $p_{i+2}$ are nearest neighbours ($\delta_{i,i+2}=1$), then remove $p_{i+1}$ from the polygon $P$ to form $P'$ if the resulting polygon $P'$ would have a smaller area than $P$.
    \item \textbf{Exchange atom:} If there is an $a_k$ such that it is a nearest neighbour of both  $p_i$  and $p_{i+2}$ ($\delta_{i,k}=\delta_{i+2,k}=1$), then replace $p_{i+1}$  with $a_k$  to form a new $P'$ if the resulting polygon $P'$ would have a smaller area than $P$.
    \item \textbf{Add new atom:} If there is an $a_k$ such that it is a nearest neighbour of both  $p_i$  and $p_{i+1}$ ($\delta_{i,k}=\delta_{i+1,k}=1$), then add $a_k$ between $p_i$  and $p_{i+1}$  to form a new $P'$ if the resulting polygon $P'$ would have a smaller area than $P$.
\end{enumerate}
If $P=P'$ the optimized pore edge has been found and the algorithm is stopped, if not make $P'$ the new $P$ and repeat the previous procedure.
\clearpage

\subsection{Pore edge atom detection}
\label{section:SI-atom-type-script}

For each possible Sulphur site exposed at the nanopore edge, a slice is made by using a  spline interpolation of the STEM image between two points perpendicular to the line connecting the two closest nanopore edge Molybdenum atoms (gray dashed lines in Fig.\ 3D). The starting point for the slice is located at a distance of two Mo-S-Mo triangle heights $2 h$ on the opposite side of the expected Sulphur atom position in respect to the Mo-Mo line. The end point of the slice is located at $2h$ away from the expected triangle site. Note that the intensity values plotted are normalized to 1 for the highest intensity pixel in the whole image, and 0 for the lowest intensity pixel.

The intensity along the slice (colored dashed lines in Fig.\ 3D) is then evaluated to determine the probability of finding a certain "atom type" at that side: a vacancy (V), double sulphurs (2S), single sulphur (S), or an invalid site (X). 

A site is declared invalid (X) if the last or fist point in the slice curve is also the maximum of the curve, unless the curve is constantly below the mid value between the expected sulphur and vacancy intensity level. In addition, if the values at the beginning or end of the slice are at the level expected for S1 or higher, they are rejected. These tests ensure that an invalid pore polygon shape is not used for further recognition. Also, each Mo polygon obtained from the pore shape optimisation algorithm can yield two possible Mo-S-Mo-$\ldots $ polygons. In order to determine the one correct polygon out of the two possibilities that match the proper crystal lattice arrangement, we perform two edge detection runs for each of the two possible polygons and reject the one with the greatest number of invalid edge sites.

The final criteria for a valid pore edge detection is the value of the most prominent peak found in the contour, as obtained using the find peaks function of the Scipy signal processing library. The minimum peak height and prominence of 0.05 and 0.01 are used, respectively. If there is no peak found or if the most prominent peak doesn't have at least twice the prominence of all the detected peaks combined, the corresponding site is declared a vacancy (V). 

If a site has not been identified as a vacancy or as invalid, the most prominent peak value is compared to the average intensities of Mo and S$_2$ obtained from the histograms of the atom sites in the entire STEM image studied (Fig.\ 3E). From this data, we obtain the probability that the site corresponds to a certain atom type, as described in the main text.

\clearpage
\begin{figure}[h!]
\centering
\includegraphics[width=0.8\linewidth]{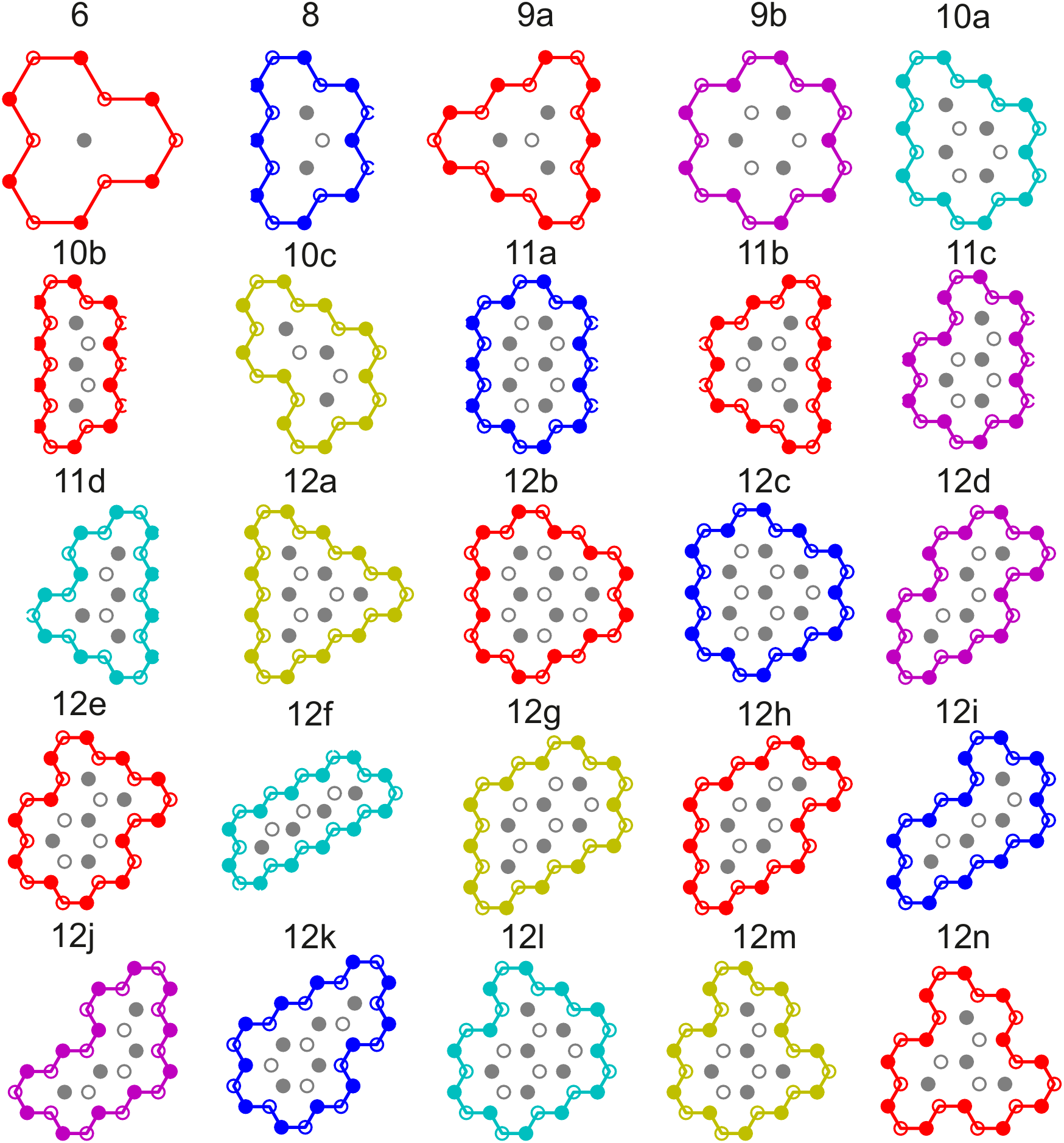}
\caption[List of detected pore shapes]{\small \textbf{Detected pore shapes}. Complete list of pores and their geometrical shapes included in the classification.}
\label{fig:SI-5-zoo}
\end{figure}

\bibliographystyle{unsrtnat}
\bibliography{refs,sasha_local}

\end{document}